\documentclass[12pt]{article}

\title{Is $G$ a conversion factor or a fundamental unit?}
\author{
G. Fiorentini\footnote{email: fiorentini@fe.infn.it} \\
INFN, Sezione di Ferrara, Ferrara, Italy \\
 L. Okun\footnote{email: okun@heron.itep.ru}, M.
Vysotsky\footnote{email: vysotsky@heron.itep.ru} \\
ITEP, Moscow, 117218, Russia}
\date{}

\begin{document}
\maketitle

\begin{abstract}

By using fundamental units $c, \hbar, G$ as conversion factors one
can easily transform the dimensions of all observables. In
particular one can make them all ``geometrical'', or
dimensionless. However this has no impact on the fact that there
are three fundamental units, $G$ being one of them. Only
experiment can tell us whether $G$ is basically fundamental.

\end{abstract}

It is well known \cite{ll} that to each mass $M$ there corresponds
a characteristic length $r_g$, the so called gravitational radius
(a body with radius $r = r_g$ forms a black hole):
$$
 r_g = 2GM/c^2 \;\; ,
$$
where $G$ is gravitational constant
$$
G = 6.673(10) \times
10^{-11} \mbox{\rm m}^3/\mbox{\rm kg s}^2 \;\; ,
$$
while $c$ is
velocity of light. Thus in all physical equations $M$ can be
substituted by $r_g$ so that mass can be ``exorcised'' from
definitions of all physical observables. As a result everything
can be measured in ``geometrical'' units of length $L$ and time
$T$ instead of standard $L$, $T$, $M$ units.

One can use $M' = GM$
instead of $M$ in order to reduce all measurements in physics to
measurements of space and time intervals and exorcise $G$ from all
equations of physics, thus reducing the number of fundamental
dimensionful constants.
We would like to make a few rather trivial remarks concerning
this proposal.

First, it is obvious that in defining $M'$  one can use $GMg(L,T)$
instead of $GM$, where $g$ is an arbitrary function of geometric
units $L, T$. In particular, in the standard case of gravitational
radius $g = 2/c^2$.

Second, as is well known (see e.g. \cite{dov}), $c, \hbar, G$ are
fundamental units in the sense that $c$ represents relativity,
$\hbar$ -- quantum mechanics, while Planck mass $m_P = \sqrt{\hbar
c/G}$ is connected with the space-time scales $l_P = \hbar/m_P c$
and $t_P = l_P/c$ at which gravity must become strong and of
quantum character. Contrary to that, the units based on $M'$
have no fundamental character.

Third, by using any of three fundamental units as a conversion
factor one does not reduce the number of fundamental units and
dimensions. E.g. when using $c$ as a unit of velocity, one
obviously preserves it as a fundamental unit. At the same time one
can measure time in units of length, or length in units of time.
However length remains length, while time remains time. Similar
considerations are valid for mass $M$, $G$ and gravitational
radius $r_g$ or any combination of the type $GMg(L,T)$.

Of course, if $G$ turns out to be only an ``effective constant''
as is the case in theories in which gravity is modified at
submillimeters distances (see e.g. \cite{vr}), then new physics
will appear well below Planck mass, maybe even at a few TeV, thus
changing the value of fundamental mass. In the case that $G$ is  only
an effective constant, a new dimensionless parameter would appear in
low-energy physics: $m_{\rm Pl}^{\rm new}/m_{\rm Pl}^{\rm old}$.
Thus the question posed in the title of this letter could be
answered by further study of the nature of gravity.

This work was partly supported by RFBR grant  No.
00-15-96562.


\begin{thebibliography}{99}
\bibitem{ll}
L.D. Landau, E.M. Lifshits, {\it The classical Theory of Fields},
1961, Oxford.
\bibitem{dov}
M. Duff, L. Okun, G. Veneziano, physics/0110060.
\bibitem{vr}
V. Rubakov, Uspekhi Fiz. Nauk {\bf 171} (September 2001) 913 (in
Russian).
\end{thebibliography}
\end{document}